\documentclass[a4paper,prb,twocolumn]{revtex4}

\usepackage{amsfonts}
\usepackage{graphicx}
\usepackage{color}
\usepackage{amsmath}
\usepackage{amssymb}
\usepackage{latexsym}
\usepackage{psfrag}

\begin{document}
\title{Effect of edge reconstruction and electron-electron interactions on quantum transport in graphene nanoribbons}
\author{S. Ihnatsenka}
\affiliation{Department of Physics, Simon Fraser University, Burnaby, British Columbia, Canada V5A 1S6}
\author{G. Kirczenow}
\affiliation{Department of Physics, Simon Fraser University, Burnaby, British Columbia, Canada V5A 1S6}

\begin{abstract}
We present numerical studies of conduction in graphene nanoribbons with reconstructed edges based on the standard tight-binding model of the graphene and the extended H\"{u}ckel model of the reconstructed defects.                    We performed atomic geometry relaxation of individual defects using density functional theory and then explicitly calculated the tight-binding parameters used to model electron transport in graphene with reconstructed edges. The calculated conductances reveal strong backscattering and electron-hole asymmetry depending on the edge and defect type. This is related to an additional defect-induced band whose wave function is poorly matched to the propagating states of the pristine ribbon. We find a transport gap to open near the Dirac point and to scale inversely with the ribbon width, similarly to what has been observed in experiments.  We predict the largest transport gap to occur for armchair edges with Stone-Wales defects, while heptagon and pentagon defects cause about equal backscattering for electrons and holes, respectively. Choosing the heptagon defect as an example, we show that although electron interactions in the Hartree approximation cause accumulation of charge carriers on the defects, surprisingly, their effect on transport is to {\em reduce} carrier backscattering by the defects and thus to enhance the conductance.
\end{abstract}

\maketitle

\section{Introduction}

The transport gaps measured experimentally in graphene nanoribbons (GNRs) exceed the theoretical electronic confinement gap.\cite{Han07, Molitor10} This discrepancy is attributed to localized states induced by edge disorder.\cite{Han07, Molitor10, Evaldsson08, Mucciolo09} Most theoretical studies of the electronic transport properties of GNRs assumed simplified edge topologies, see e.g., Refs. \onlinecite{Evaldsson08, Mucciolo09, disorder09}. However several experimental studies have characterized individual edge defects by means of Raman spectroscopy, scanning tunneling microscopy or transmission electron microscopy.\cite{Casiraghi09, Kobayashi06, ZLiu09, Girit09, Koskinen09} The high degree of chemical reactivity of graphene edges favors edge reconstruction with different topologies, some of which were examined in recent {\em ab initio} calculations.\cite{Koskinen08, Song10, Huang09} Dubois \textit{et al}.\cite{Dubois10} showed that realistic edge topologies strongly affect electron transport in the armchair GNRs: The conductance can be suppressed by orders of magnitude for either electrons or holes, depending on the defect geometry.\cite{Dubois10} Heptagon defects were found to act as donors and to suppress the electron conductance much more strongly than the hole conductance whereas pentagon defects were found to act as acceptors and to suppress the hole conductance much more strongly than the electron conductance.\cite{Dubois10} More recently, a few defect topologies for armchair and zigzag GNRs were studied theoretically in Ref. \onlinecite{Hawkins12}. In contrast to Ref. \onlinecite{Dubois10}, these authors \cite{Hawkins12} reported pentagon and heptagon edge reconstructions to have little effect on the ribbon conductance near the Dirac point.\cite{difference} The origin of this difference between the predictions of these two studies\cite{Dubois10, Hawkins12} remains unclear. Furthermore, theoretical estimates of the {\em transport} gaps in GNRs are still not available for realistic models of edge reconstruction, and the implications for the interpretation of experimental data\cite{Han07, Molitor10} are yet to be explored. It has also been shown that electron-electron interactions modify electron conduction in GNRs substantially.\cite{ee, hartree12} Specifically, when the electron Fermi level is not at the charge neutrality point, electrons are predicted to accumulate along the edges of the ribbon and therefore any edge imperfection may be expected to play an important role in transport. However, the interplay between this charge redistribution effect and edge reconstruction in the context of transport is yet to be examined either theoretically or experimentally. Thus, the detailed understanding of electron transport in GNRs with realistic topological edge defects is still incomplete. 

In this paper we report systematic studies of the electronic transport in realistic edge-disordered GNRs with both zigzag and armchair edges (zGNR and aGNR). Three different defect topologies are examined: pentagon, heptagon and Stone-Wales. The Stone-Wales mechanism reconstructs zGNR edges into alternating pentagon-heptagon pairs (z57), while in aGNR it causes two separate `armrest' hexagons to merge into adjacent heptagons (a757). To describe realistic defect topologies we first relax the atomic geometries using density functional theory.\cite{gaussian} We then obtain the tight-binding (TB) parameters for the relaxed geometries from the standard quantum chemical parameterization of the extended H\"{u}ckel model.\cite{yaehmop,Kirczenow} 

We note that the previous theoretical studies of transport in GNRs with reconstructed topological defects\cite{Dubois10, Hawkins12} employed TB parameters obtained by fitting tight binding models to the results of density functional theory (DFT) based calculations in different ways. As we have already mentioned, those transport calculations\cite{Dubois10, Hawkins12} yielded qualitatively different results. For this reason, and because density functional theory-based transport calculations have serious fundamental limitations that often render their results unreliable,\cite{Kirczenow} we chose to base our TB parameters instead on the extended H\"{u}ckel model whose parameters were derived from a large body of experimental molecular electronic structure data.\cite{yaehmop} We incorporate these TB parameters into the standard TB Hamiltonian of graphene. As will be discussed below, the transport results that we obtained using this TB model agree reasonably well with those in Ref. \onlinecite{Dubois10} for the systems that were studied in Ref. \onlinecite{Dubois10}. 

We identify the defect bands in the electronic band structures of the GNRs with reconstructed edge defects and relate strong electron backscattering in these GNRs to the presence of these bands. We estimate realistic transport gaps for the reconstructed GNRs with randomly located defects and find gap sizes to depend on the type of defect and to be ordered in the following way: a757 $>$ a7 $\gtrsim$ a5 $>$ z57 $>$ z5 $\gtrsim$ z7. We find all of the transport gaps to scale inversely with ribbon width, consistent with experimental findings.\cite{Han07, Molitor10} 

Finally we report on calculations with electron-electron interactions taken into account in the Hartree approximation for a representative class of systems with single and multiple heptagon defects in aGNRs. Consistent with previous studies,\cite{ee, hartree12} we find the electron-electron interactions to give rise to charge redistribution towards the edges of the ribbons when gating shifts the ribbon Fermi energy away from the Dirac point. This results in enhanced concentrations of the charge carriers on the defects at the edges of the ribbon. Naively, one might expect this to result in increased backscattering of the charge carriers by the edge defects and therefore a decrease of the ribbon conductance due to electron-electron interactions. However, surprisingly, we find the opposite. Namely, we find the conductances of ribbons with these edge defects to {\em increase} for most values of the Fermi energy when electron-electron interactions are introduced into the model. We explain this counterintuitive behavior as arising from to spatial rearrangement of the low energy electronic eigenstates of the ribbons and screening of local density of states fluctuations, both of which result from electron-electron interactions. This weakening of the effects of edge disorder on transport  due to electron-electron interactions is a novel effect that appears to be unique to graphene devices. It may have played a role in the conductance quantization that has been observed in recent experiments\cite {Lin08,Lian2010,Tombros11} on graphene nanostructures despite the presence of edge imperfections. 

\section{Model}
\label{model}

We consider suspended graphene nanoribbons, adopting a similar approach to that in Ref. \onlinecite{hartree12}. The GNR is separated from the back gate by dielectric and air layers and is attached at its two ends to semiinfinite leads represented by ideal ribbons having the same width $W$. The system is described by the Hamiltonian 
\begin{equation}
  H = \sum_i V^H_i\;a_i^{\dag }a_i - \sum_{\left\langle i,j\right\rangle} t_{ij}\left( a_i^{\dag}a_j + h.c. \right),
  \label{eq:hamiltonian}
\end{equation}
where $t_{ij}=t=2.7$ eV is the matrix element between nearest-neighbor atoms; $V^H_i$ is the Hartree potential at atom $i$ which results from the Coulomb interaction with the uncompensated charge density $-en$ in the system (including the image charges). In coordinate space the Hartree potential can be written as
\begin{equation}
  V^H(\mathbf{r})=\frac{e^{2}}{4\pi \varepsilon _{0}\varepsilon}\int d\mathbf{r}\,^{\prime } \sum_k \frac{n_k(\mathbf{r}^{\prime})}{\sqrt{|\mathbf{r}-\mathbf{r}^{\prime}|^{2}+b_k^{2}}}  ,
  \label{eq:V_H}
\end{equation}
where $-en_k(\mathbf{r}^{\prime})$ is the $k^{th}$ electron or image charge placed at distance $b_k$ from the graphene layer. The integration in \eqref{eq:V_H} was performed over the whole device including the semiinfinite leads. For details of our model we refer the reader to \onlinecite{hartree12}. Note that spin effects are outside of the scope of present study. 

\begin{figure*}[t]
\includegraphics[keepaspectratio,width=\textwidth]{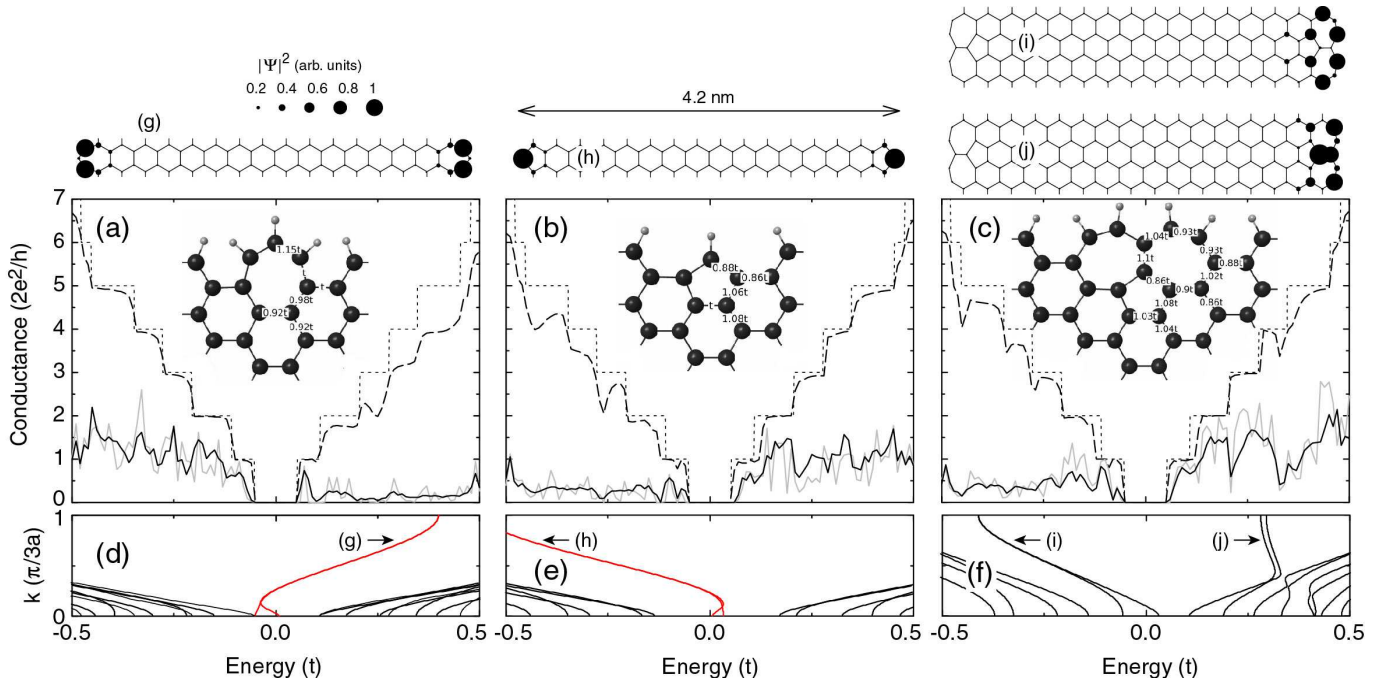}
\caption{The conductances (a)-(c) and defect band structures (d)-(f) for armchair graphene nanoribbons with heptagon, pentagon and Stone-Wales edge reconstruction defects, respectively. The respective single-defect geometries along with the tight-binding parameters are illustrated in the insets of (a)-(c). The conductances are presented for the ideal ribbon (dotted line), the ribbon with a single defect (dashed line) and 30 randomly distributed defects (solid gray line). The solid black line in (a)-(c) corresponds to the averaged conductance over 100 configurations. The band structures in (d)-(f) are calculated for the defect periodically repeated along both edges of the ribbon. The bands due to defects (shown in red in (d) and (e)) are clearly distinct from the standard dispersion of the armchair ribbon and correspond to the Fermi energies where strong electron or hole backscattering occurs. The top plots (g)-(j) show the square modulus of the wave function for the energies marked by the arrows in (d)-(f). All ribbons are 4.2 nm wide and 500 nm long, which corresponds to 33 carbon atoms in the cross section and 1173 unit cells along the device.}
\label{fig:1}
\end{figure*}

\begin{figure*}[th]
\includegraphics[keepaspectratio,width=\textwidth]{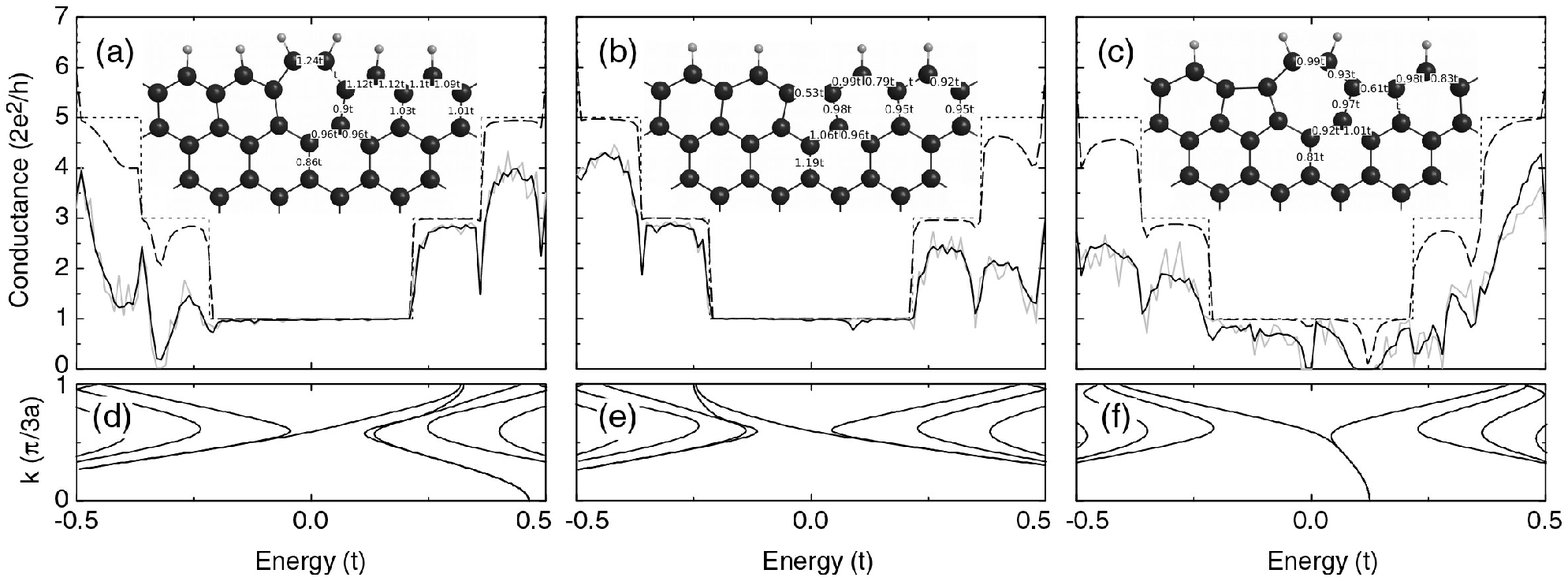}
\caption{The same as Figs. \ref{fig:1}(a)-(f) but for zigzag graphene ribbons. The ribbon dimensions $W=4.2$ nm and $L=500$ nm correspond to 20 carbon atoms in the cross section and 2033 unit cells along the device.}
\label{fig:2}
\end{figure*}

The edge reconstruction is introduced as randomly located defects at the edges of the hexagonal graphene lattice. We suppose low concentrations of defects that do not cluster though later we shall comment on such a case. We performed geometry relaxations for the edge reconstruction in GNRs using the Gaussian 09 software package.\cite{gaussian} The structures studied were graphene half-disks of several tens of carbon atoms passivated at the edges with hydrogen, the defects being near the center of the straight edge. The carbon atoms belonging to defect core as well as several nearest atoms at the edge were allowed to relax freely, the other carbon atoms being held fixed in the standard hexagonal graphene geometry with the C-C distance of 1.42 \AA. The whole structure was kept planar. The relaxed structures obtained in this way are shown in the insets in the Figures \ref{fig:1} and \ref{fig:2}. The matrix elements in the Hamiltonian Eq. (\ref{eq:hamiltonian}) were modified to account for reconstructed topology by calculating the relevant matrix elements within the extended H\"{u}ckel model.\cite{yaehmop,Kirczenow} For further details and comparison with tight binding parameterizations estimated from density functional theory we refer the reader to our earlier study of adsorbed atoms and molecules on graphene.\cite{adsorbate} The modified matrix elements are given in the insets in the Figures \ref{fig:1} and \ref{fig:2}.

Having calculated the reconstructed geometries and corresponding Hamiltonian matrix elements we find the conductance 
\begin{equation}
	G = -\frac{2e^{2}}{h} \int dE \; \sum_{ij} T_{ij}(E) \frac{\partial f(E-E_F)}{\partial E}
	\label{eq:conductance}
\end{equation}
as a function of the Fermi energy $E_F$. Here $T_{ij}(E)$ is the transmission coefficient from subband $j$ in the left lead to the subband $i$ in the right lead, at energy $E$ and $f$ is the Fermi function. $T_{ij}(E)$ is calculated by the recursive Green's function method.\cite{Xu08}

\section{Results}

We investigate the transport properties of edge reconstructed GNRs of realistic sizes, similar to those investigated experimentally.\cite{Han07, Molitor10} Two host/edge graphene orientations are studied in the following: armchair and zizgag. The edges are assumed to be randomly decorated by either heptagon, pentagon or joint heptagon-pentagon (Stone-Wales) defects. 

Before we proceed with these edge reconstructions we shall comment on the effect of edge relaxations that preserve the benzenoid topology of pristine armchair and zigzag graphene edges. In this case, the outermost carbon atoms, when relaxed, move toward graphene host which in turn leads to incremental changes of the hopping energies between the outermost carbon atoms. In the armchair configuration, we find the Hamiltonian matrix elements $t_{ij}$ between the two outer atoms of the armrest to increase by 9$\%$ while the hopping matrix elements between these atoms and their other nearest neighbors are enhanced by 3$\%$. In the zigzag configuration, the two nearest neighbor hopping energies $t_{ij}$ to/from the outermost carbon atoms increase by 3$\%$. With these marginal changes incorporated, the transport properties stay qualitatively unchanged although the electronic subband bottoms shift by few percent. 

Because any topological edge defect causes much more dramatic changes in the electron transport properties we shall in what follows neglect the above hexagon relaxation except in the vicinity of the topological defects. For the sake of clarity we shall begin by considering electron transport within the non-interacting approach, setting $V^H=0$ in Eq. \ref{eq:hamiltonian}.

\subsection{Armchair ribbons}

Figure \ref{fig:1}(a) shows the conductances of aGNRs with heptagon edge defect(s). The relaxed defect geometry along with the estimated hopping energies between neighboring carbon atoms are given in the inset. The conductance for a single relaxed heptagon (the dashed line in Figure \ref{fig:1}(a)) develops stronger backscattering for higher electron subbands. As the defect concentration grows, the conductance becomes suppressed more strongly with pronounced oscillations due to quantum wave interference. After averaging over different defect configurations, conductance dips due to enhanced backscattering near subband bottoms become apparent, as has been discussed previously in context of simple atomic lattice defects.\cite{disorder09} This property appears to be generic for any edge reconstruction defect. For sufficiently large defect concentrations the electron transport mechanism undergoes a transition from ballistic transport to Anderson localization in these 500nm long GNRs. 

The strong electron-hole conductance asymmetry in Fig. \ref{fig:1}(a) for heptagon defects randomly distributed along the edges of the ribbon (solid grey and solid black lines) is qualitatively similar to that predicted by Dubois \textit{et al}. \cite{Dubois10} for armchair ribbons with heptagon defects. They found heptagon defects to act as donors and quasi-bound states associated with them to result in strong electron scattering and depressed conductances at positive energies.\cite{Dubois10} We also find that if the heptagon defect is repeated periodically along the edges of the ribbon an additional electron subband appears in the GNR spectrum mainly at positive energies; this is the subband shown in red in Fig. \ref{fig:1}(d). The associated Bloch states are exponentially\cite{Park13} localized at the edges of the ribbon as is shown in Fig. \ref{fig:1}(g). We note that this additional subband is poorly matched to the Bloch states propagating in the pristine GNR leads and therefore results in strong electron backscattering in the positive energy range. For heptagon defects randomly distributed along the GNR's edges this additional electron subband breaks up into electron defect states that also mediate electron backscattering, giving rise to the much stronger suppression of electron conductance than hole conductance in Fig. \ref{fig:1}(a).
(Note that the band structure in Fig. \ref{fig:1}(d) is an idealized one presented only for the purpose of the present discussion since it is computed using the tight-binding parameters of the core of the isolated defect; for extended 
or clustered defects some of the tight-binding parameters will differ.)

By contrast, pentagon edge defects (see the inset in Fig. \ref{fig:1}(b)) in armchair ribbons have been predicted to behave as acceptors.\cite{Dubois10} Consistent with this we find them to introduce hole states, or in the case of pentagons periodically repeated along the edges, an additional hole subband shown in red in Fig. \ref{fig:1}(e). The result is strong hole backscattering and therefore much stronger suppression of the hole conductance than of the election conductance, see Fig. \ref{fig:1}(b), as in Ref. \onlinecite{Dubois10}. 

The effect of the Stone-Wales defect (Fig.\ref{fig:1}(c)) is more subtle because it doesn't change number of carbon atoms at the ribbon's edge. For periodically repeated Stone-Wales defects the aGNR unit cell is doubled and the band structure exhibits both electron and hole bands associated with the defect; see Figs \ref{fig:1}(f),(i),(j). The Bloch states are localized on different graphene sublattices (Figs \ref{fig:1}(i),(j)) and cause differing backscattering. We found somewhat stronger backscattering for holes than for electrons for this 757 defect, see Fig. \ref{fig:1}(c), whereas Ref. \onlinecite{Dubois10} predicted similar scattering for electrons and holes, slightly stronger for the electrons.

\subsection{Zigzag ribbons}

The conductances and band structures for GNRs of nominally the same sizes and having the same numbers of reconstructed topological defects as in Figure \ref{fig:1} but for the zigzag edge configuration are shown in Figure \ref{fig:2}. Some significant differences between the zigzag and armchair cases are as follows. Firstly, the scattering of the carriers by the defects is generally weaker for the zigzag case. Secondly, the z575 defect causes especially strong conductance suppression near the flat band of the structure with periodically repeated defects; see Fig. \ref{fig:2}(c),(f). Thirdly, the flat band of the pristine zGNR is strongly affected and no longer present near the graphene Dirac point for heptagon and pentagon defects but not so for the z575 defect. The strong modification of the subband structure in the first two cases results from the topological properties of the zigzag edge because it supports charge localization at the edges where defects form. The much weaker perturbation of the flat band for the z575 defect may be explained by effective narrowing of the ribbon such that the zigzag edge moves effectively into the ribbon interior.

\begin{figure}[th]
\includegraphics[keepaspectratio,width=0.8\columnwidth]{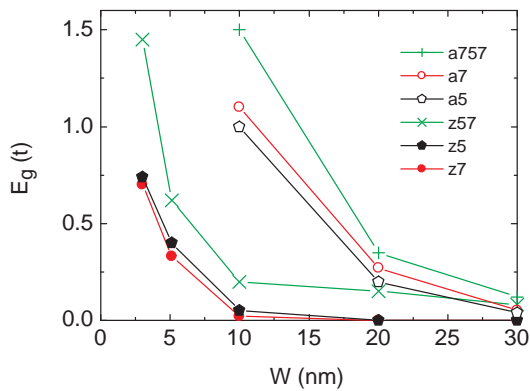}
\caption{(Color online) The transport gap $E_g$ of GNRs with 440 randomly placed edge reconstruction defects as a function of ribbon width. The length is fixed at $L=1$ $\mu$m. The widths of the aGNRs are chosen so that conduction at the Dirac point is metallic for the defect-free ribbons.} 
\label{fig:3}
\end{figure}

\subsection{Transport gaps}

The GNRs studied above in Figs. \ref{fig:1} and \ref{fig:2} were only 4.2 nm wide and 500 nm long, but the analysis presented applies equally well to other geometries including realistic sizes fabricated experimentally.\cite{Han07, Molitor10} An important property of disordered ribbons measured in the experiments is the transport gap, $E_g$. This quantifies the energy (or gate voltage) window where the conductance is suppressed below some threshold value. Figure \ref{fig:3} shows the calculated $E_g$ as a function of the ribbon width. The length of ribbon is fixed at $L=1$ $\mu$m and the number of defects is 440. The threshold value was chosen to be half of the conductance quantum. We found that, for the reconstructed GNRs with randomly located defects, $E_g$ scales inversely with the ribbon width.  This agrees qualitatively with experimental data.\cite{Han07, Molitor10} We find the magnitudes of $E_g$ to be ordered as follows  a757 $>$ a7 $\gtrsim$ a5 $>$ z57 $>$ z5$\gtrsim$ z7 according to the type of defect and ribbon edge. The heptagon and pentagon defects cause approximately the same gap to develop. Note that this width dependence of of $E_g$ is very different than that for bulk defects where $E_g$ was found to be independent of $W$.\cite{adsorbate}

\begin{figure}[th]
\includegraphics[keepaspectratio,width=\columnwidth]{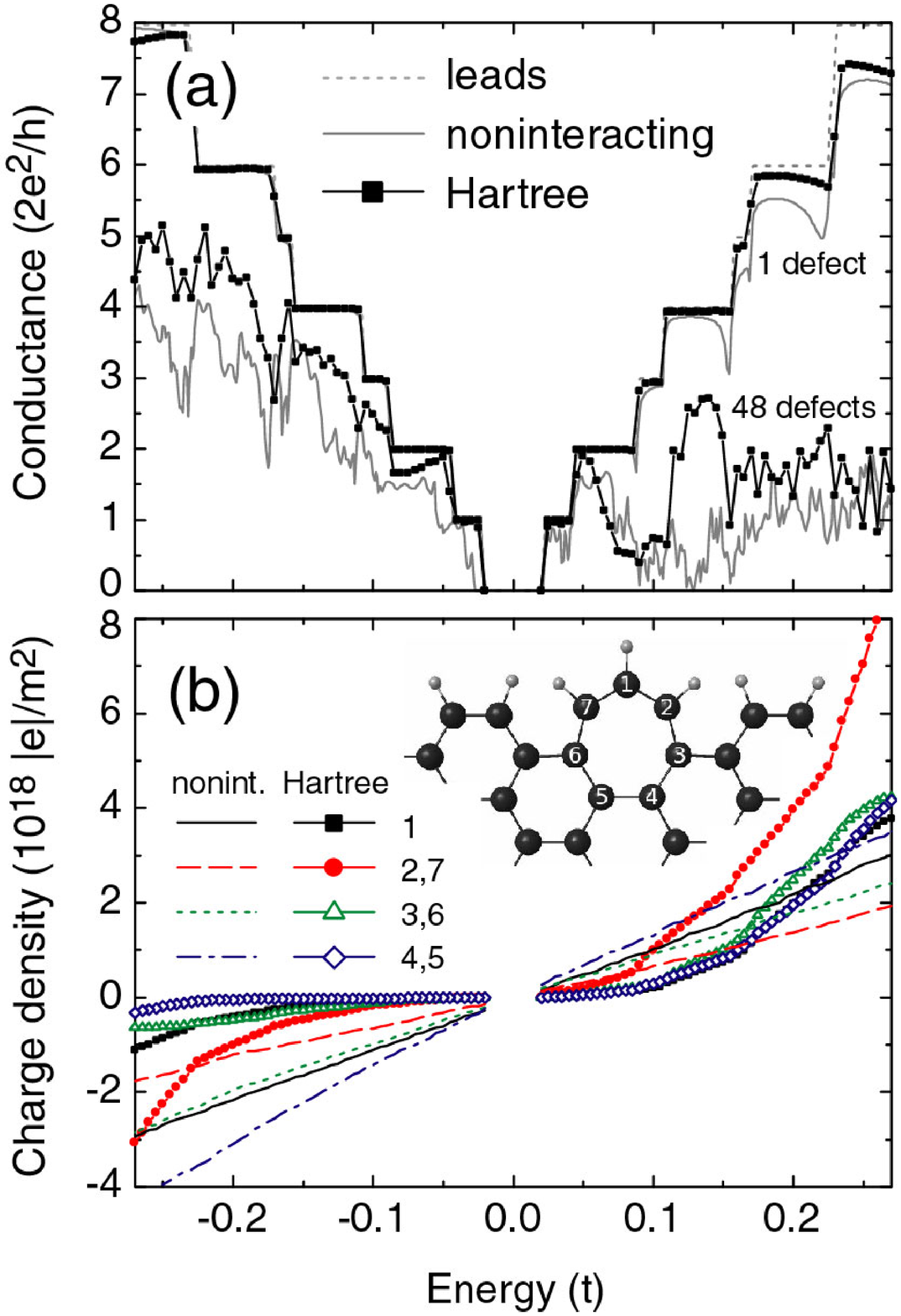}
\caption{(Color online) (a) The conductances of aGNRs with heptagon defects calculated in the non-interacting and Hartree approaches. 48 heptagon defects corresponds to $10\%$ of the hexagon rings being replaced. (b) The charge densities  on individual carbon atoms of the heptagon defect as numbered in the inset. $L=100$ nm, $W=10$ nm, temperature $T=10$ K.} 
\label{fig:4}
\end{figure}

\subsection{Electron-electron interactions}

Having studied the effects due to edge reconstruction within the non-interacting electron model, we are in a position to consider the effects of electron interactions. However, an approach that accounts for Coulomb interactions as formulated in the section \ref{model} demands huge computational resources that prohibit the study of GNRs of realistic sizes. Therefore we will restrict further study to a representative aGNR of $W=10$ nm and $L=100$ nm with heptagon defects only and will focus on the new physics introduced by the electron interactions. Figure \ref{fig:4} shows the conductance and charge densities on individual carbon atoms as a function of the Fermi energy. 
Previous theoretical work\cite{ee, hartree12} has shown electron-electron interactions to result in redistribution of charge towards the ribbon edges when the ribbon is gated away from the charge neutrality point. This effect is responsible for the non-linear growth of the calculated charge densities on the atoms of the heptagon edge defect in the Hartree approximation as the Fermi energy moves away from the Dirac point in Figure \ref{fig:4}(b). The stronger charge accumulation at positive than negative energies in the Hartree approximation in Figure \ref{fig:4}(b) is due to the donor character of the heptagon edge defect. In stark contrast, the non-interacting model for the same single defect location predicts linear charge accumulation on the defect  in Figure \ref{fig:4}(b) and slightly larger hole densities than electron densities. 

Thus electron-electron interactions result in the redistribution of charge towards the ribbon edges where the edge defects are located. However, surprisingly, we find that this does {\em not} result in stronger electron backscattering by the edge defects in the Hartree approximation than in the non-interacting electron model. To the contrary, we find the net effect of the electron-electron interactions to be, in most cases, to {\em weaken} electron backscattering by the edge defects and therefore to {\em increase} the ribbon conductance. 

For a single edge defect, the weakening of electron backscattering due to electron-electron interactions manifests itself most clearly at positive energies in Figure \ref{fig:4}(a) where it results in weakening or complete suppression (in the Hartree approximation) of  the conductance dips at subband edges that are a prominent feature of the non-interacting model. For 48 heptagon defects distributed randomly along the edges of the ribbon, as can be seen in Figure \ref{fig:4}(a), the calculated conductance in the Hartree approximation is substantially higher than that for the non-interacting electron model for most values of the Fermi energy.

We attribute this counterintuitive behavior to a combination of two physical mechanisms, namely, (i) spatial rearrangement of the low energy electronic eigenstates due to the formation of triangular electrostatic potential wells at the edges of the sample and (ii) screening of local density of states features associated with electron backscattering. Mechanism (i) is as follows: If electron-electron interactions are included in the model, the excess electronic charge on the ribbon gives rise to triangular electrostatic potential wells at the edges of the ribbon. Each low energy electron eigenstate then localizes in one of these wells or the other. Thus, in the presence of electron-electron interactions, electrons in the low energy eigenstates backscatter only from defects on one edge of the ribbon; defects at the other edge have little effect on them. Thus the low energy electrons only interact with  approximately half of the defects in the sample in the model with electron-electron interactions, although they interact with each of these defects somewhat more strongly than in the absence of the electron-electron interactions. Of these two competing effects, the reduction in the number of defects with which the electron interacts has the larger effect on the conductance than does the stronger coupling. Therefore the conductance increases when electron-electron interactions are turned on. The screening mechanism (ii) is more subtle. Local charge fluctuations associated with electronic scattering resonances are screened or anti-screened due to electron-electron interactions. The effect of this on electron backscattering is strongly energy-dependent: At energies near subband edges the screening tends to strongly suppress charge fluctuations and the associated electron backscattering is also suppressed. At other energies screening weakly enhances the charge fluctuations, however, the conductance is also enhanced. Thus mechanism (i) appears to have a larger effect on the conductance than does mechanism (ii) at those energies.
   
\section{Discussion}

Our results demonstrate strong charge scattering by the edge reconstruction defects, the scattering strength being different for electron and holes, for both armchair and zigzag ribbons. This results in electron-hole asymmetry in the conductances of GNRs with such defects. We find the transport gap in the GNR conductance to scale inversely with the ribbon width, suggesting that the edge reconstruction defects may be responsible for the gaps observed in experiments\cite{Han07, Molitor10}. The gaps were fitted by $E_g=\alpha/(W-W^{\ast})$ with $W^{\ast}=16$ nm and $\alpha=0.2$ eV nm and $\alpha=0.38$ eV nm in Refs. \onlinecite{Han07} and \onlinecite{Molitor10}, respectively. $W^{\ast}$ accounts for inactive edges whose widths are not those of the defect regions themselves but rather are determined by the extent of the disorder-induced localized surface-type states.\cite{Evaldsson08} Fitting the above formula to our band gaps in Fig. \ref{fig:3} yields $\alpha=0.04...0.8$ eV nm and $W^{\ast}=2...10$ nm depending on edge and defect type. These values agree reasonably well with experimental findings\cite{Han07, Molitor10} though the absolute values of $E_g$ exceed the experimental values for most of the defects except z5 and z7. It is plausible that the defects studied here were not the only defects in experiments, but might be accompanied by other types of the defects. 

Our study for aGNRs supports the results and conclusions drawn by Dubois etal.\cite{Dubois10} but not of Hawkins etal.\cite{Hawkins12}. The latter claimed the low bias conduction of reconstructed GNRs to be similar to the that of the pristine GNRs. By contrast, we found strong electron scattering due to reconstructed edge defects near $E=0$. It should be emphasized that our study, for first time, to the best of our knowledge, provides simple and realistic tight-binding parameters for edge reconstruction defects that we explicitly calculated in the extended H\"{u}ckel model. Strong electron and/or hole backscattering is interpreted as being due to mismatch between propagating states in the pristine GNRs and the states associated with the defects. 

The influence of edge reconstruction defects on the transport properties might be understood from the Clar sextet theory.\cite{Dubois10, Baldoni08, Clar} This proposes an interpretation of the electronic properties in terms of valence bond theory based on the localization of aromatic sextets. According to Clar's theory, zGNRs can be viewed as the superposition of Kekul\'{e} structures with a finite number of benzene rings in each row of hexagons.\cite{Baldoni08} For aGNRs, the Clar representation depends on ribbon width, being either fully benzenoid, Kekul\'{e} or incomplete Clar structures. Upon the introduction of any edge reconstruction defects, the bonding can be considered as the superposition of two mirroring Kekul\'{e} structures that partially destroy the benzenoid character of GNRs. By increasing the localization of $\pi$ electrons in carbon-carbon double bonds, such defects destroy the local aromaticity at the ribbon edge and are thus expected to have large impact on $\pi$-$\pi^{\ast}$ conduction channels.\cite{Dubois10}

We studied isolated edge reconstruction defects that are assumed to be surrounded by the pristine hexagonal graphene lattice. This picture is valid for low defect concentrations when neighboring defects do not merge. However, the model that we have presented might be further extended to clustered groups of defects that were, for example, observed in some experiments.\cite{Chuvilin09} In this case, neighboring defects will affect the geometry of the relaxed structure and the tight-binding parameters $t_{ij}$ in the Hamiltonian Eq. \ref{eq:hamiltonian} will be different. We have performed a test calculation for periodic Stone-Wales defects in a zGNR and found tight-binding parameters close to those reported in the literature.\cite{Hawkins12, Rakyta10} For those parameters the band structure, that is similar to Fig. \ref{fig:2}(f), reveals downward bending of the flat band, i.e. the flat band occurs below the graphene Dirac point, in agreement with $ab$ $initio$ band structures in the literature.\cite{Koskinen08, Song10, z57band, Nguyen11}

It is worth noting that the present formulation of the model does not include next neighbor hopping in the Hamiltonian \ref{eq:hamiltonian}. We estimate that hopping to be $0.27t$ for pristine graphene. While inclusion of this parameter will substantially complicate the calculation we expect that it would not change the results presented above qualitatively. It is also worth pointing out that, in realistic devices, the edge reconstruction regions are not infinitely long but rather are sandwiched between electron reservoirs. This is equivalent to the transport problem we formulated and studied here whereas the band gaps obtained for infinitely long periodic defect structures\cite{Hawkins12, Nguyen11} do not straightforwardly corresponds to the transport gaps in such devices.

In the case where ribbon edges contain several different defects the combined effect on transport appears to be roughly additive.\cite{Dubois10} For example, equal concentrations of heptagon and pentagon defects restore electron-hole conductance symmetry and act in concert to increase the transport gap value.

Based on our findings, electron-electron interactions play a significant role in transport in graphene ribbons. Although they result in redistribution of charge to the edges of the ribbon where the edge defects are located,
their net effect on transport is to reduce the backscattering of charge carriers by edge defects and thus to increase the ribbon conductance. While electron-electron interactions affect the conductance quantitatively, at the Hartree level of approximation they do not change the transport properties of the ribbon considered in the present study qualitatively. For example, the electron-hole asymmetry of the conductance in Fig. 4 is maintained. 

\section{Conclusion}

Our quantum transport calculations have shown edge reconstruction to cause strong electron/hole scattering and transport gaps to occur in GNRs in agreement with the experiments of Han etal.\cite{Han07} and Molitor etal.\cite{Molitor10} Strong scattering is a manifestation of mismatch between propagating states in the pristine ribbon and the states due to the defect. It is accompanied by electron-hole conductance asymmetry for defects that behave as donors or acceptors. We show that electron-electron interactions at the Hartree level of approximation result in charge accumulation at edge defects when the Fermi energy (or the gate voltage) varies but predict weaker electron backscattering by the defects in comparison to the noninteracting approach. Our results suggest that edge reconstruction may affect electron transport strongly in many experimentally realized graphene based devices. Work is in progress to address this issue in the context of the 0.7 anomaly in graphene nanoconstrictions observed recently by Tombros etal.\cite{Tombros11}


\begin{thebibliography}{99}

\bibitem{Han07} M. Y. Han, B. \"{O}zyilmaz, Y. Zhang, and P. Kim, Phys. Rev. Lett. \textbf{98}, 206805 (2007).
\bibitem{Molitor10} F. Molitor, C. Stampfer, J. Guttinger, A. Jacobsen, T. Ihn and K. Ensslin, Semicond. Sci. Technol. \textbf{25} 034002 (2010).
\bibitem{Evaldsson08} M. Evaldsson, I. V. Zozoulenko, Hengyi Xu and T. Heinzel, Phys. Rev. B \textbf{78}, 161407(R) (2008).
\bibitem{Mucciolo09} E. R. Mucciolo, A. H. Castro Neto, and C. H. Lewenkopf, Phys. Rev. B \textbf{79}, 075407 (2009).
\bibitem{disorder09} S. Ihnatsenka and G. Kirczenow, Phys. Rev. B \textbf{80}, 201407(R) (2009).
\bibitem{Casiraghi09} C. Casiraghi, A. Hartschuh, H. Qian, S. Piscanec, C. Georgi, A. Fasoli, K. S. Novoselov, D. M. Basko and A. C. Ferrari, Nano Lett. \textbf{9}, 1433 (2009).
\bibitem{Kobayashi06} Y. Kobayashi, K. I. Fukui, T. Enoki, K. Kusakabe, Phys. Rev. B \textbf{73} 125415 (2006).
\bibitem{ZLiu09} Zheng Liu, Kazu Suenaga, P. J. F. Harris and S. Iijima, Phys. Rev. Lett. \textbf {102}, 015501 (2009).
\bibitem{Girit09} \c{C}. \"{O}. Girit, J. C. Meyer, R. Erni, M. D. Rossell, C. Kisielowski, Li Yang, Cheol-Hwan Park, M. F. Crommie, M. L. Cohen, S. G. Louie, A. Zettl, Science \textbf{323}, 1705 (2009).
\bibitem{Koskinen09} P. Koskinen, S. Malola, and H. H\"{a}kkinen, Phys. Rev. B \textbf{80} 073401 (2009).
\bibitem{Koskinen08} P. Koskinen, S. Malola, and H. H\"{a}kkinen, Phys. Rev. Lett. \textbf{101} 115502 (2008).
\bibitem{Song10} L. L. Song, X. H. Zheng, R. L. Wang, and Z. Zeng, J. Phys. Chem. C \textbf{114}, 12145 (2010).
\bibitem{Huang09} Bing Huang, Miao Liu, Ninghai Su, Jian Wu, Wenhui Duan, Bing-lin Gu, and Feng Liu, Phys. Rev. Lett. \textbf{102}, 166404 (2009).
\bibitem{Dubois10} S. M.-M. Dubois, A. Lopez-Bezanilla, A. Cresti, F. Triozon, B. Biel, J.-C. Charlier, and S. Roche, ACS Nano \textbf{4}, 1971 (2010).
\bibitem{Hawkins12} P. Hawkins, M. Begliarbekov, M. Zivkovic, S. Strauf, C. P. Search, J. Phys. Chem. C \textbf{116}, 18382 (2012).
\bibitem{difference}See, for example, Fig. 3 of Ref. \onlinecite{Hawkins12}.
\bibitem{ee} J. Fern\'{a}ndez-Rossier, J. J. Palacios, L. Brey, Phys. Rev. B \textbf{75}, 205441 (2007); P. G. Silvestrov and K. B. Efetov, Phys. Rev. B \textbf{77}, 155436 (2008); A. A. Shylau, J. W. Klos, and I.V. Zozoulenko, Phys. Rev. B \textbf{80}, 205402 (2009); D. I. Odili, Y. Wu, P. A. Childs, and D. C. Herbert, J. Appl. Phys. \textbf{106}, 024509 (2009); D. A. Areshkin and Branislav K. Nikoli\'{c}, Phys. Rev. B \textbf{81}, 155450 (2010); J. J. Palacios, J. Fern\'{a}ndez-Rossier, L Brey and H A Fertig, Semicond. Sci. Technol. \textbf{25}, 033003 (2010); A. A. Shylau, I. V. Zozoulenko, H. Xu and T. Heinzel, Phys. Rev. B \textbf{82}, 121410(R) (2010); F. T. Vasko and I. V. Zozoulenko, Appl. Phys. Lett. \textbf{97}, 092115 (2010); Z. Wang and R. W. Scharstein, Chem. Phys. Lett. \textbf{489}, 229 (2010); M. V. Medvedyeva and Ya. M. Blanter, Phys. Rev. B \textbf{83}, 045426 (2011); A. A. Shylau and I. V. Zozoulenko, Phys. Rev. B \textbf{84}, 075407 (2011).

\bibitem{hartree12} S. Ihnatsenka and G. Kirczenow, Phys. Rev. B \textbf{86}, 075448 (2012).

\bibitem{gaussian} M. J. Frisch \textit{et al.}, computer code GAUSSIAN 03, revision A.02 (Gaussian Inc., Pittsburgh, PA, 2009).
The HSEh1PBE hybrid functional and  6-311G(d) basis set were used in the geometry relaxations carried out in the present
study.

\bibitem{yaehmop}The version of extended H\"{u}ckel theory used 
was that of J. H. Ammeter, H.-B. B\"{u}rgi, J. C. Thibeault, and R. 
Hoffman, J. Am. Chem. Soc. {\bf 100}, 3686 (1978) as implemented in
the YAEHMOP numerical package by G. A. Landrum and W. V. Glassey
(Source-Forge, Fremont, California, 2001).
\bibitem{Kirczenow}  For a recent review see G. Kirczenow, \textit{Molecular nanowires and their properties as electrical conductors}, The Oxford Handbook of Nanoscience and Technology,Volume I: Basic Aspects, Chapter 4, edited by A. V. Narlikar and Y. Y. Fu, Oxford University Press, U.K. (2010).
\bibitem{Lin08} Yu-Ming Lin, V. Perebeinos, Zhihong Chen, and Ph. Avouris, Phys. Rev. B \textbf{78}, 161409(R) (2008).
\bibitem{Lian2010} C. Lian, K. Tahy, T. Fang, G. Li, H. G. Xing, and D. Jena, Appl. Phys. Lett. {\bf 96}, 103109 (2010).
\bibitem{Tombros11} N. Tombros, A. Veligura, J. Junesch, M. H. D. Guimaraes, I. J. Vera-Marun, H. T. Jonkman and B. J. van Wees, Nature Physics \textbf{7}, 697 (2011).
\bibitem{Rakita10} P. Rakyta, A. Kormanyos, J. Cserti, and P. Koskinen, Phys. Rev. B \textbf{81}, 115411 (2010)
\bibitem{adsorbate} S. Ihnatsenka and G. Kirczenow, Phys. Rev. B \textbf{83}, 245442 (2011).
\bibitem{Xu08} H. Xu, T. Heinzel, M. Evaldsson, and I. V. Zozoulenko, Phys. Rev. B \textbf{77}, 245401 (2008).
\bibitem{Park13} C. Park, J. Ihm, G. Kim, arXiv:1304.0314
\bibitem{Baldoni08} M. Baldoni, A. Sgamellotti, F. Mercuri, Chem. Phys. Lett. \textbf{464}, 202 (2008).
\bibitem{Clar} E. Clar, \textit{The Aromatic Sextet}, Wiley, London (1972).
\bibitem{Rakyta10} P. Rakyta, A. Korm\'{a}nyos, J. Cserti, and P. Koskinen, Phys. Rev. B \textbf{81}, 115411 (2010).
\bibitem{z57band} J. N. B. Rodrigues, P. A. D. Goncalves, N. F. G. Rodrigues, R. M. Ribeiro, J. M. B. Lopes dos Santos, and N. M. R. Peres, Phys. Rev. B \textbf{84}, 155435 (2011)
\bibitem{Nguyen11} L. Tung Nguyen, C.Huy Pham, V. Lien Nguyen, J. Phys. Condens. Matter \textbf{23}, 295503 (2011)
\bibitem{Chuvilin09} A. Chuvilin, J. Meyer, G. Algara-Siller, and U. Kaiser, New J. Phys. \textbf{11}, 083019 (2009)


\end{thebibliography}
\end{document}